\documentclass{article}

\usepackage{gensymb}
\usepackage{PRIMEarxiv}

\usepackage[utf8]{inputenc} 
\usepackage[T1]{fontenc}    
\usepackage{hyperref}       
\usepackage{url}            
\usepackage{booktabs}       
\usepackage{amsfonts}       
\usepackage{nicefrac}       
\usepackage{microtype}      
\usepackage{lipsum}
\usepackage{fancyhdr}       
\usepackage{graphicx}       
\graphicspath{{media/}}     

\title{Sputter-deposited PZT-on-Silicon Optimization for C-band Electro-Optic Modulation}

\author{
  Suraj, Shankar Kumar Selvaraja \\
  Center for Nanoscience and Engineering \\
  Indian Institute of Science, Bangalore 560012, India \\
  \texttt{\{Suraj,Shankar Kumar Selvaraja\} suraj2@iisc.ac.in, shankarks@iisc.ac.in} \\
}

\begin{document}
\maketitle

\begin{abstract}
Ferroelectric materials exhibit interesting electric, mechanical and optical properties. Particularly, tin-film lead zirconium titanate (PZT) has been used as a standard piezo-electric material in micro-electro-mechanical systems. Interestingly, it has one of the highest electro-optic properties that can be exploited to make high-speed on-chip light modulators. In this work, we demonstrate sputter-deposited PZT on silicon-on-insulator wafers. Desired phase and film morphology is achieved through process and buffer layer engineering. We present results on buffer layer screening between MgO, TiO$_2$ and Pt. MgO is identified as a suitable buffer layer from an optical and film growth perspective. We achieve a measured coercive field for PZT/MgO and PZT/Pt film is 50 kV/cm and 30 kV/cm, respectively. We use a PZT-loaded ring modulator with an electro-optic response of  14 pm/V.
\end{abstract}


\section{Introduction}
The need to modulate light at a higher frequency for various applications has increased. Notably, the roadmap for communication calls for increased network bandwidth and reduced power consumption. Due to well-established material processing and fabrication technologies, Silicon (Si) photonic integrated circuits offer a scalable platform to address some of the requirements. However, the charge carrier modulation on electric field application for modulators is limited at a high frequency of operation (limited by the transients).\textsuperscript{\cite{Multi-photon,chen2023review}} Recently, ferroelectric materials have been investigated to bring the gap in performance without compromising the scalability offered by Si photonics. Perovskites with ABO$_3$ empirical such as Lithium Niobate (LNO), Barium Titanate (BTO), and Lead Zirconium Titanate (PZT) are being considered for high frequency, optically lossless modulation and non-linear photonic applications.\textsuperscript{\cite{opt4010003, Perspective, Ban:22, BaTiO3}} Perovskites such as LNO, BTO, and PZT have properties such as a high Pockels coefficient, making them the choice of material to achieve pure phase modulation-based high-frequency electro-optic (EO) modulators with no material losses. Among them, PZT offers an advantage over LNO due to large Pockels coefficient along with higher spontaneous polarization which could be harnessed in non-linear photonic applications apart from being lossless at 1500 nm.\textsuperscript{\cite{du1998crystal,feutmba2020strong,spirin1998measurement,uchiyama2007electro,haertling1972recent,chen1989anionic, SAHOO2016299, NUFFER20003783}}
With such properties, PZT is an efficient choice for high-speed EO modulators and non-linear optic devices. Despite the immense advantage offered by PZT over Si photonics, little work has been done on on-chip sputter deposited PZT-based photonics or integrating sputtered PZT on Si for photonic devices.\textsuperscript{\cite{singh2021sputter, ZHU201922324}}

Deposition methods such as chemical solution deposition(CSD), evaporation, molecular beam epitaxy, metal-organic chemical vapor deposition (MOCVD), pulse laser deposition (PLD), and sputter deposition are used to deposit PZT film.\textsuperscript{\cite{sakashita1991preparation,das2001preparation,izyumskaya2006growth}} Most of the reported works on PZT-based modulators have used a sol-gel-based or aerosol-based deposited PZT .\textsuperscript{\cite{alexander2018nanophotonic,busch1992linear,nakada2009lanthanum,feutmba2019hybrid,lee1999drying,swartz1991sol,ramamurthi1992electrical,teowee1995electro}} CSD is limited by the thickness of the film that can be deposited in one step and the wet process makes it prone to contamination. Evaporation, MBE, and MOCVD depend on the vapor pressure of the component of PZT and hence makes it difficult to obtain a stoichiometric film.\textsuperscript{\cite{muralt2006texture,fushimi1967phase}} Furthermore, the carbon contamination associated with an organo-metallic precursor in MOCVD and MBE is also challenging.
The high substrate temperature and a different rate of reaction of components of PZT in MBE and MOCVD would require a continuous change in the flow rates to compensate for the change in stoichiometry.\textsuperscript{\cite{itoh1991mocvd}} PLD and sputter can be used to get a higher deposition rate compared to MBE, which has a very low deposition rate. Sputter has the advantage of wafer-scale deposition limited by target size compared to deposition in PLD, which is constrained by plume size and scalability.\textsuperscript{\cite{zhang2019flexible,abe1991pzt,singh2021sputter}} 

This work presents a detailed growth optimization of RF sputtered PZT-on-Si. The desired quality of PZT is achieved through buffer layer selection and process optimization. We explored MgO, TiO$_2$, and Pt as growth buffer layers. The growth and annealing process is optimized for the fabrication of crack density mitigation for the photonic device. We demonstrate a PZT-on-SOI electro-optic modulator with a 14 pm/V DC response.

\section{PZT growth optimization}
\begin{figure}[htbp]
\centering\includegraphics[width=\linewidth]{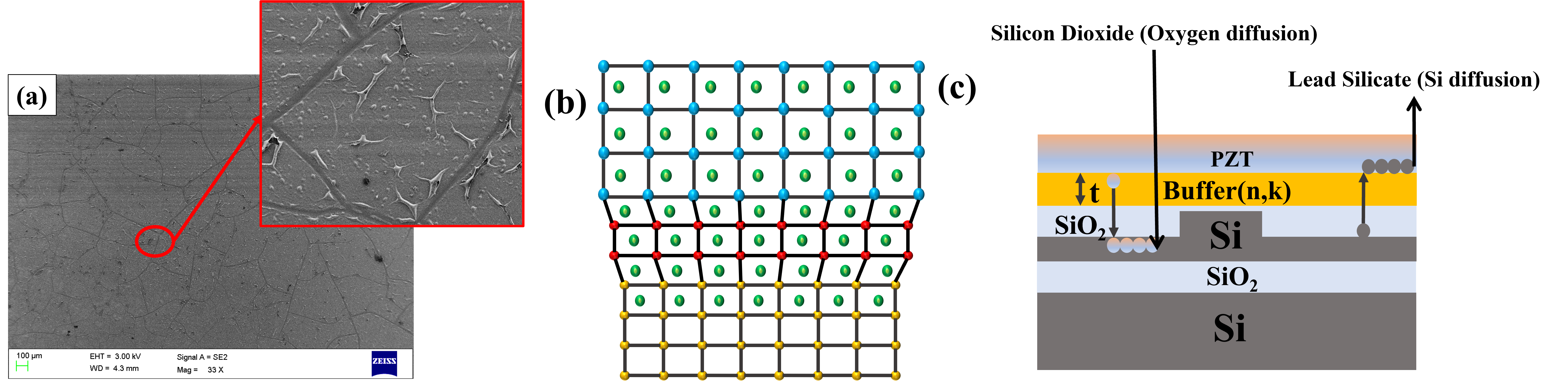}
\caption{(a) SEM image of PZT deposited directly on Si; (b) general schematic of lattice matching required to obtain an epitaxial, crack-free film; (c) proposed structure for PZT film deposition optimization for photonic device fabrication.}
\label{fig:PZT-On-Si}
\end{figure}
Lattice constant mismatch of $\approx$25 \% between Si and PZT would result in stress-induced cracks in the deposited PZT film on Si substrate. The field emission scanning electron microscope (FESEM) image in \textbf{Figure \ref{fig:PZT-On-Si}(a)} confirms the crack formation in PZT film deposited on Si. Such cracked PZT films are a source of scattering losses in the photonic devices upon integration and lead to an increase in the leakage current that is detrimental to EO devices. Therefore, it is essential to minimize the lattice mismatch to reduce the crack density to obtain low scattering loss and a smooth film. \textbf{Figure \ref{fig:PZT-On-Si}(b)} shows the generalized schematic of the transition steps in lattice parameter, which reduces the stress developed at the interface, making a crack-free epitaxial or highly oriented film deposition possible with \textbf{Figure \ref{fig:PZT-On-Si}(c)} showing the proposed cross-section schematic of the stack for photonic application.
\begin{table}[]
\centering
\caption{Buffer layer lattice constant and refractive index }
\label{tab:Lattice parameter and Refractive Index}
\begin{tabular}
{|c|c|c|c|}
\hline
Sl. No & Material & lattice constant(nm)(a) &  Refractive Index (at 1550 nm) \\
\hline
1     & Si      & 0.54301  & 3.43           \\
\hline
2     & MgO      & 0.4216 & 1.65    
\\
\hline
3     & PZT  & 0.404 & 2.3 \\
\hline
4     & TiO$_2$  & 0.38  & 2.45   
\\
\hline
5     & Pt  & 0.392   &   
\\
\hline
\end{tabular}
\end{table}

A thin buffer layer increases the interaction of the optical field with the PZT layer that is required for EO modulators. The issue with a thin buffer layer of thickness less than $\approx$ 20 nm is an inefficient diffusion barrier for Si, Pb, and O atoms. Lead silicate formation in PZT (Si diffusion towards PZT) and SiO$_2$ formation (O diffusion from PZT to Si) at Si/buffer interface leads to higher annealing temperature (diffusion barrier reduces the activation energy for perovskite formation) for perovskite phase formation, crack density as well as the presence of pyrochlore phase in the film due to Pb deficiency in PZT film.\textsuperscript{\cite{kim1997influence,basit1998growth,song1997effects,yong2000effects,choi2015investigation,willems1995influence,kukushkin2012mechanism,peng2013improvement}} SiO$_2$ thus serves the purpose of aiding the buffer layer as a diffusion barrier, reducing the annealing temperature required to obtain perovskite phase for application in electro-optic modulators.

\subsection{Buffer layer optimization}

\begin{figure}[htbp]
\centering\includegraphics[width=\linewidth]{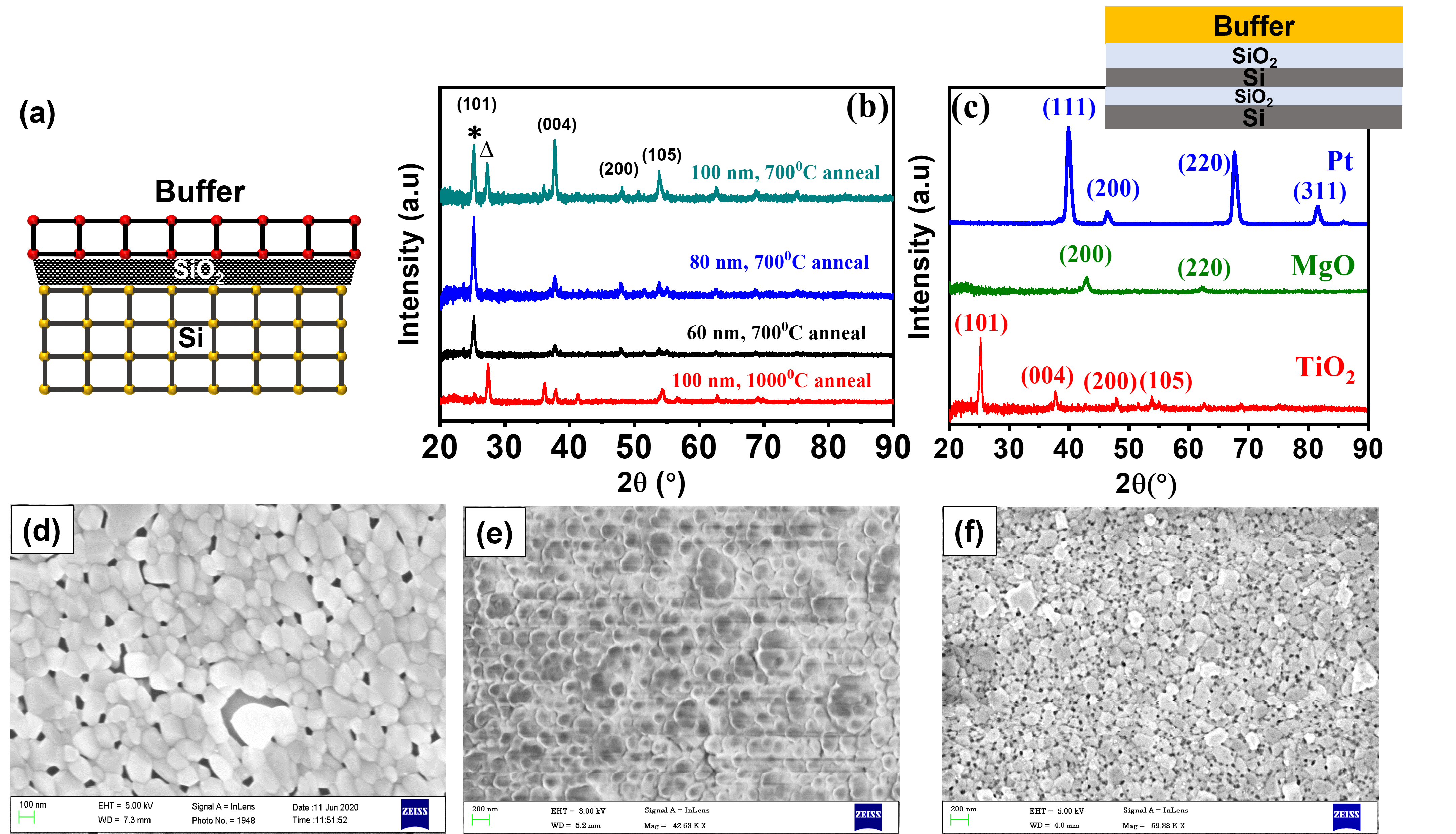}
\caption{(a) Schematic for amorphous buffer layer/SiO$_2$/Si stack; XRD spectra for (b) anatase and rutile TiO$_{2} $ with "*" corresponding to anatase phase and "$\triangle$" representing rutile phase, (c) anatase TiO$_2$, Pt and MgO; FESEM image of (d) TiO$_{2} $; (e) MgO and (f) Pt/Ti.}
\label{fig:Buffer_layer}
\end{figure}

\textbf{Figure \ref{fig:Buffer_layer}(a)} shows the stack that is used to deposit buffer on Si with intermediate SiO$_{2} $ layer. Based on \textbf{Table \ref{tab:Lattice parameter and Refractive Index}}, due to small lattice mismatch with PZT, Pt and 1550 nm transparent MgO, TiO$_2$ were used as buffers for PZT film deposition and characterization.\textsuperscript{\cite{gilmore2003growth,tonejc2002analysis,yano1994epitaxial}} Sputtering was used to deposit the buffer layer on the SiO$_2$/Si stack with a 300 nm SiO$_2$ layer. The parameters used to sputter deposit MgO and TiO$_2$ are RF power of 100 W and source to target distance of 10 cm. Ti/Pt was deposited with a DC power of 150 W/35 W for a 20 nm and 100 nm thickness, respectively. TiO$_2$ thickness of 60 nm, 80 nm, and 100 nm was annealed at 700\degree C and 1000\degree C respectively to verify the transition from anatase to rutile. MgO thickness of 80 nm was annealed at 700\degree C. All anneal were performed in an air ambiance. As seen in \textbf{Figure \ref{fig:Buffer_layer}(b)}, polycrystalline TiO$_{2} $ with anatase phase is formed for thickness less than 100 nm for annealing temperature of 700\degree C. For a thickness of 100 nm at an annealing temperature of 700\degree C, both rutile and anatase phase is observed, which on further increasing anneal temperature to 1000\degree C, forms a rutile phase. To use a buffer in a photonic device, we need a thin film and a lower temperature to achieve the crystalline phase, making the anatase TiO$_2$ a preferable choice.\textsuperscript{\cite{sreemany2009influence}} \textbf{Figure \ref{fig:Buffer_layer}(c)} shows consolidated XRD spectra of polycrystalline film of anatase TiO$_{2} $, MgO and Pt on SiO$_{2} $/Si stack. TiO$_2$ is observed to be predominantly (101) oriented, Pt as highly oriented towards (111) plane while MgO giving a (200) as a preferred orientation. FESEM images of deposited and annealed TiO$_{2} $, MgO, Pt on SiO2/Si with an average surface roughness of $\approx$ 8 nm, 2 nm, and 5 nm respectively are shown in \textbf{Figure \ref{fig:Buffer_layer}(d),(e)\&(f)} confirming a continuous film. 
\begin{figure}[t]
\centering\includegraphics[width=0.9\linewidth]{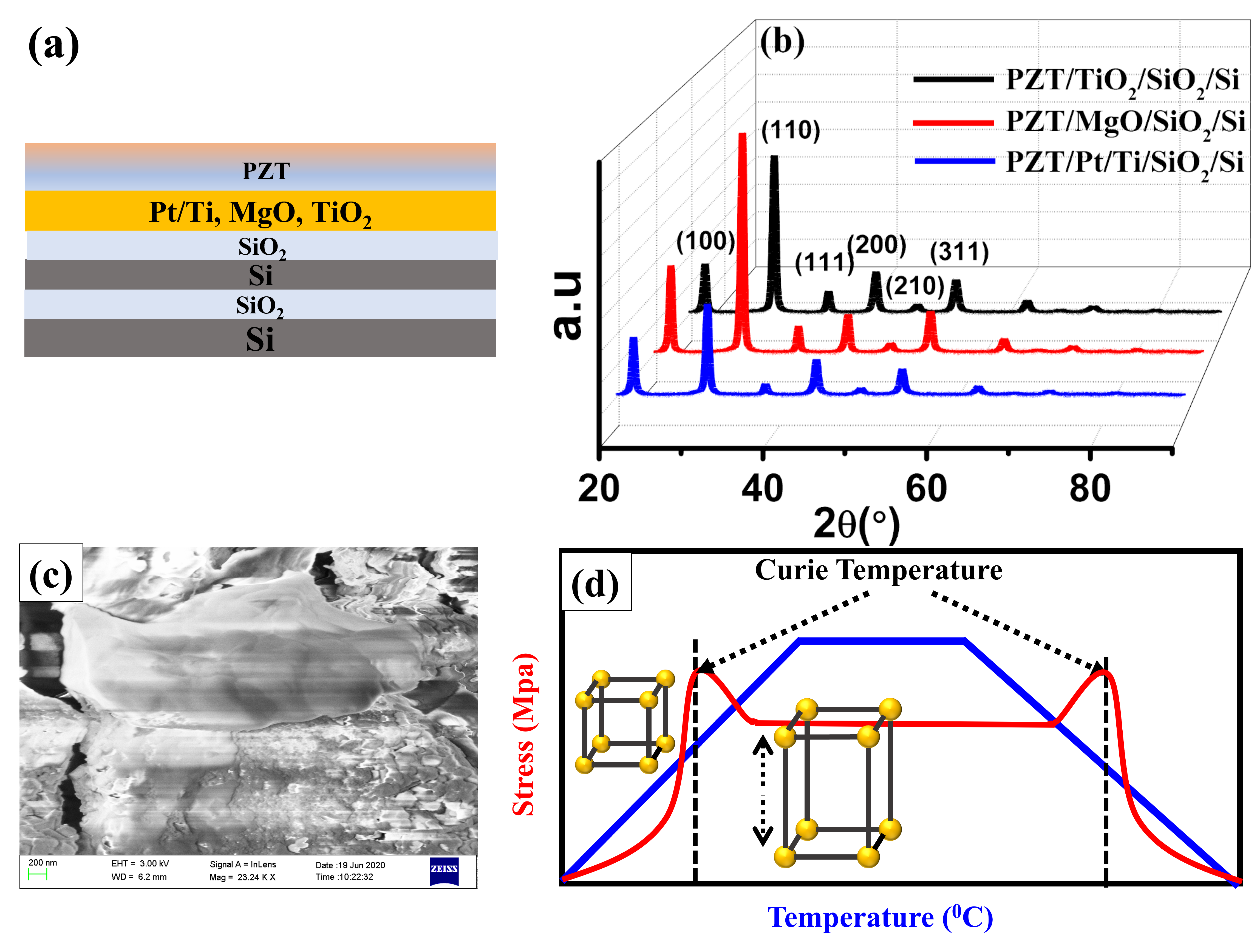}
\caption{(a) Schematic of PZT deposited stack; (b) XRD of PZT on anatase TiO$_{2} $, MgO and Pt on  SiO$_{2} $/Si stack; (c) FESEM showing cracks present in deposited PZT at fast anneal rate(20\degree C/min and a dwell temperature of 650\degree C); (d) proposed variation of stress with annealing temperature.}
\label{fig:PZT_Deposition}
\end{figure}

\subsection{Crack density optimization}

\textbf{Figure \ref{fig:PZT_Deposition}(a)} shows the PZT deposited stack with Pt/SiO$_2$, MgO/SiO$_2$ and TiO$_2$/SiO$_2$ layer as a buffer layer on silicon-on-insulator (SOI) stack. Current-Voltage (I-V), Capacitance-Voltage (C-V), Polarization-Electric field (P-E), and piezo-electric coefficient determination would be performed on the stack for application in opto-MEMS and electro-optic device fabrication using PZT as an active material. Ti/Pt buffer layer deposition thickness is 120 nm, with Ti being 20 nm and Pt being 100 nm, while for TiO$_{2} $ and MgO, the thickness was 80 nm. PZT was sputter deposited on TiO$_2$, MgO, and Pt with RF power of 100 W, a source-target distance of 4 cm, argon flow of 30 sccm with an ex-situ anneal temperature of 650 \degree C and a ramp rate of 21 \degree C/min. The thickness of PZT obtained is 1$\mu$m. The film was air annealed at 650 \degree C  with a dwell time of 1 hr and ramp time of 30 min. \textbf{Figure \ref{fig:PZT_Deposition}(b)} shows the XRD spectra for the post-annealed polycrystalline PZT with dominant PZT XRD peak of (100) and (110) for MgO, TiO$_{2} $ and Pt buffer layer. FESEM image in  \textbf{Figure \ref{fig:PZT_Deposition}(c)} shows cracks present in the deposited and ex-situ annealed PZT film with a ramp of 20 \degree C/min and a dwell temperature of 650 \degree C. Critical parameter that affects the crack density in PZT deposition are dwell temperature and ramp rate of the PZT during the annealing process.
The formation of cracks can be explained by \textbf{Figure \ref{fig:PZT_Deposition}(d)} wherein there is a sudden change in the stress developed in the deposited PZT film at curie temperature due to a change in the lattice structure at morphotropic phase boundary (MPB) for Pb(Zr$_{0.58} $Ti$_{0.42} $)O$_{3} $ leading to crack generation.  

\begin{figure}[htbp]
\centering\includegraphics[width=\linewidth]{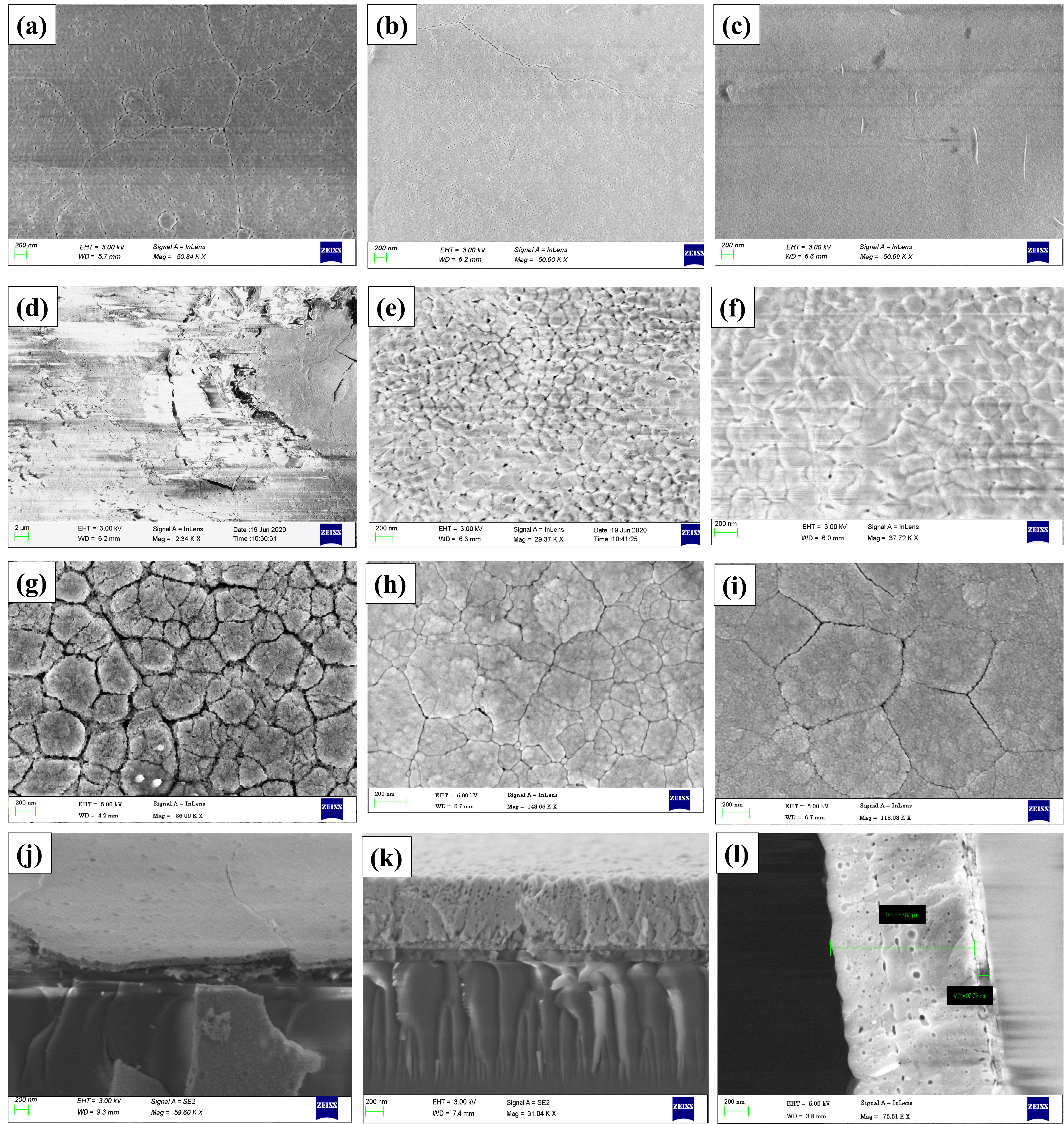}
\caption{FESEM images of PZT on MgO annealed with a ramp rate of 5\degree C/min at (a) 650\degree C,(b) 550\degree C anneal and (c) 550\degree C anneal at 3.6\degree C/min.FESEM image of PZT on TiO$_2$ annealed with a ramp rate of 5\degree C/min at (d) 650\degree C,(e) 550\degree C anneal, and (f) at 550\degree C anneal at 3.6\degree C/min.FESEM image of PZT on Pt annealed with a ramp rate of 5\degree C/min at (g) 650\degree C,(h) 550\degree C anneal, and (i) at 550\degree C anneal at 3.6\degree C/min.(j)FESEM image of Crack density in PZT at 20\degree C/min, Crack free film at 3.6\degree C/min for (k) MgO and (l) Pt.}
\label{fig:SEM_optimized}
\end{figure}

PZT film was deposited at an RF power of 100 W with a sample-to-target distance of 4.2 cm. Ex-situ annealing was done at 650 \degree C and 550 \degree C at a ramp rate of 5 \degree C/min 3.6 \degree C/min. FESEM images of PZT annealed at a ramp rate of 5\degree C/min and dwell temperature of 650\degree C shows more cracks as compared to PZT film annealed at 550\degree C as seen in \textbf{Figure \ref{fig:SEM_optimized}(a)} and (b). \textbf{Figure \ref{fig:SEM_optimized}(c)} and (d) show the reduction in the crack density for MgO and TiO$_2$ buffers. The corresponding FESEM images of crack reduction in TiO$_2$ and Pt, for annealing at 650\degree C with 20\degree C/min ramp rate, 550 \degree C with 5 \degree C/min and 3.6 \degree C/min ramp rate, are shown in \textbf{Figure \ref{fig:SEM_optimized}(d-i)}. 
Surface roughness measurement using atomic force microscopy was done for the optimized surfaces giving a roughness of $\approx$2 nm and $\approx$ 10 nm for MgO and TiO$_2$ film, respectively, while surface roughness obtained for a Pt buffer is about 5 nm. From the optimization summary presented in \textbf{Figure \ref{fig:SEM_optimized}} MgO and Pt buffered PZT growth yields dense and low crack density that is desired for photonic applications. Reduced ramp rate led to a crack-free film with no pin-holes as is evident in the cross-section SEM image of a cracked and optimized MgO film shown in \textbf{Figure \ref{fig:SEM_optimized}(j\&k)} with the optimized PZT/Pt cross-section shown in \textbf{Figure \ref{fig:SEM_optimized}(l)}. The obtained PZT film thickness post-anneal was $\approx$900 nm.

The piezo-electric and optical characteristics of the deposited film are characterized using Pt and mgO buffered layers, respectively. Since MgO is non-conducting, it is used to characterize lateral leakage, polarization and electro-optic response. Pt is used to characterize vertical leakage and polarization characteristics.

\subsection{Electrical characterization}
\begin{figure}[htbp]
\centering\includegraphics[width=\linewidth]{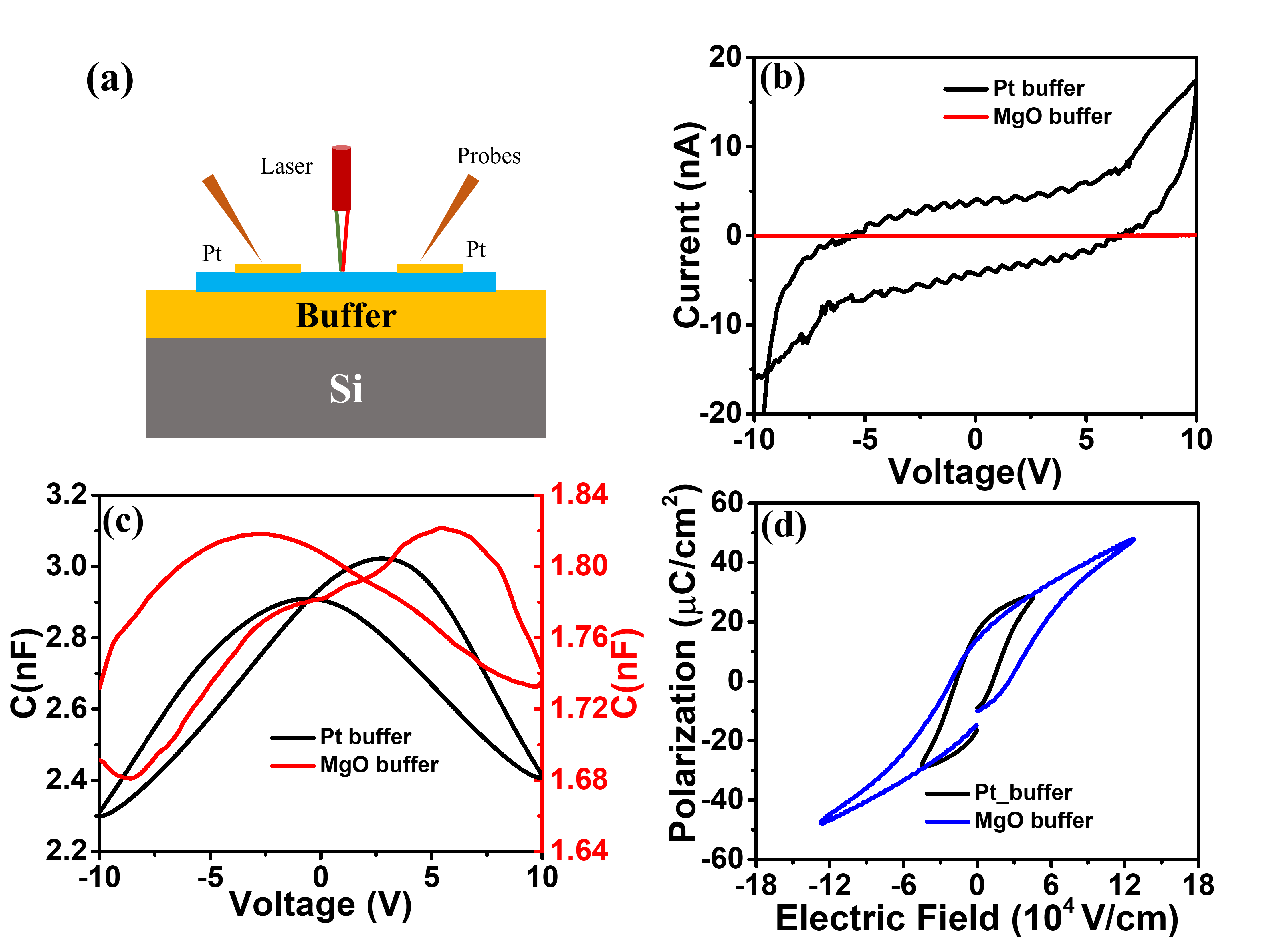}
\caption{(a) Schematic of the device fabricated for testing film quality using co-planar electrodes with Pt and MgO buffered PZT ;(b) I-V characterization of PZT film ; (c) C-V characterization of PZT film; (d) P-E loop characterization of PZT film.}
\label{fig:Functional_char}
\end{figure}

\begin{figure}[t]
\centering\includegraphics[width=\linewidth]{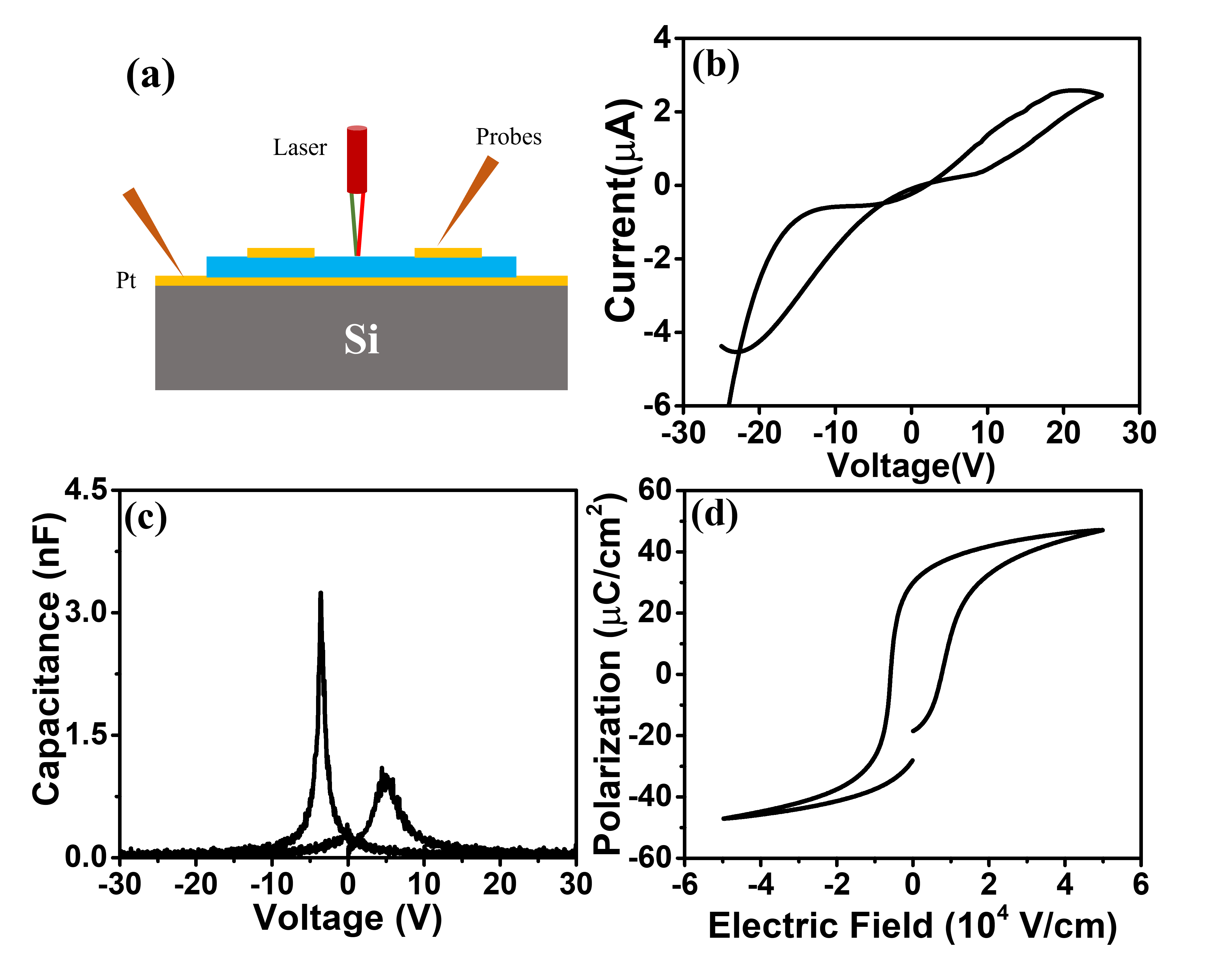}
\caption{(a) Schematic of the device fabricated for testing film quality using top-bottom electrodes with Pt buffered PZT;(b) I-V characteristic of the PZT film; (c) C-V characterization and; (d) P-E loop characterization of PZT film with the top-bottom electrode.}
\label{fig:Pt_bottom_electrode}
\end{figure}

\textbf{Figure \ref{fig:Functional_char}(a)} shows the cross-section schematic of the device used for electrical characterization of the PZT film using a co-planar Pt electrode on PZT film buffered on Pt and MgO to qualify the piezo-electric property for use in an EO device. An optimized parameter of 100 W RF power, source-to-target distance of 4.2 cm, anneal temperature of 550 \degree C, and ramp rate of 3.6 \degree C/ min was used to deposit PZT of thickness 0.9 $\mu$m. The bottom Pt/Ti electrode used was 20 nm/100 nm in thickness, and the top electrode of 20 nm/80 nm in thickness. The bottom electrode was annealed at 650 \degree C for 40 min.

A Peak current of $\approx$20 nA for Pt buffer and $\approx$7 pA for MgO buffer film (\textbf{Figure \ref{fig:Functional_char}(b)} with co-planar electrode. The low current shows that there is no leakage path between electrodes on PZT. The co-planar electrodes could be used for poling and electric field modulation.

\textbf{Figure \ref{fig:Functional_char}(c)} shows the C-V curve for PZT/Pt and PZT/MgO for co-planar electrode while \textbf{Figure \ref{fig:Functional_char}(d)} shows the P-E loop for the two configurations. Piezoelectric characteristic is evident from the dual peak in capacitance with a maximum capacitance of 3 nF and 1.9 nF and a polarization hysteresis with a maximum value of $\approx$40 $\mu$C/cm$^2$ and $\approx$60 $\mu$C/cm$^2$ for Pt and MgO buffer layers, respectively. The electrodes were defined by Pt pad size of 1X1 mm$^{2}$, a gap of 7 $\mu$m and a PZT thickness of 900 nm. The coercive field calculated for Pt buffered and MgO buffered PZT is 30 kV/cm and 50 kV/cm, respectively. Schematically shown in \textbf{Figure \ref{fig:Pt_bottom_electrode}(a)}, the top-bottom electrode, is used to confirm the presence of cracks in the film that could lead to electric short. The vertical electrode configuration characterization of the PZT/Pt layer as shown in \textbf{Figure \ref{fig:Pt_bottom_electrode}(a-c)} confirms the crack-free nature of the film as well as the piezo-electric behavior of the PZT film with a coercive field of $\approx$12 kV/cm.

\section{Device fabrication}
\begin{figure}[htbp]
\centering\includegraphics[width=0.9\linewidth]{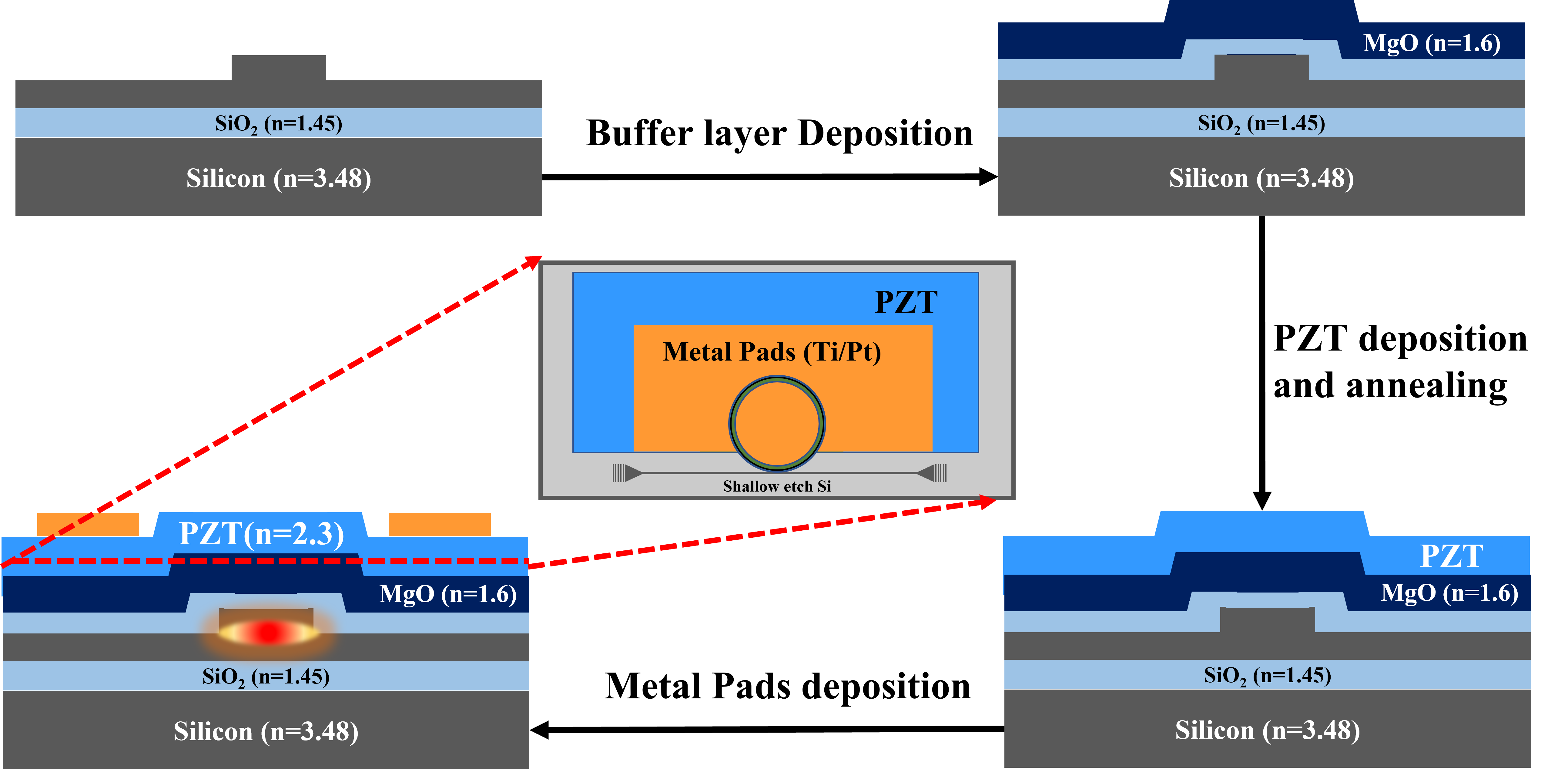}
\caption{Process flow to fabricate an EO modulator loaded with sputtered PZT.}
\label{fig:Process_flow}
\end{figure}

\textbf{Figure \ref{fig:Process_flow}} shows the flow diagram of the EO device fabrication. Shallow etch Si micro-ring resonators were fabricated on an SOI wafer with a 200 nm Si on top of a 2000 nm buried SiO$_{2}$ layer with 70 nm etch depth and 600 nm waveguide width. Electron beam lithography was used to pattern the waveguide using a negative photoresist and dry etch to pattern a micro-ring resonator in the SOI substrate. A deposition of 100 nm of 
SiO$_{2} $ was done followed by 80 nm and 1 $\mu$m of MgO and PZT deposition, respectively. An optimized deposition parameter for MgO used was RF power of 100 W, a source-to-target distance of 10 cm, and an annealing temperature of 700\degree C. The corresponding values for PZT used were RF power of 100 W, a source-to-target distance of 4.2 cm, and an annealing temperature of 550\degree C with a ramp rate of 3.6\degree C/min. The etching of PZT was done using a wet etch solution of H$_{2}$O, BHF (buffered HF), HCl, and H$_{2}$O$_{2}$ in the ratio 10:1:1:1. Co-planar Ti/Pt electrode of thickness 20 nm/100 nm were deposited using electron beam evaporator on the PZT layer.

\section{Electro-Optic characterization}

\begin{figure}[htbp]
\centering\includegraphics[width=\linewidth]{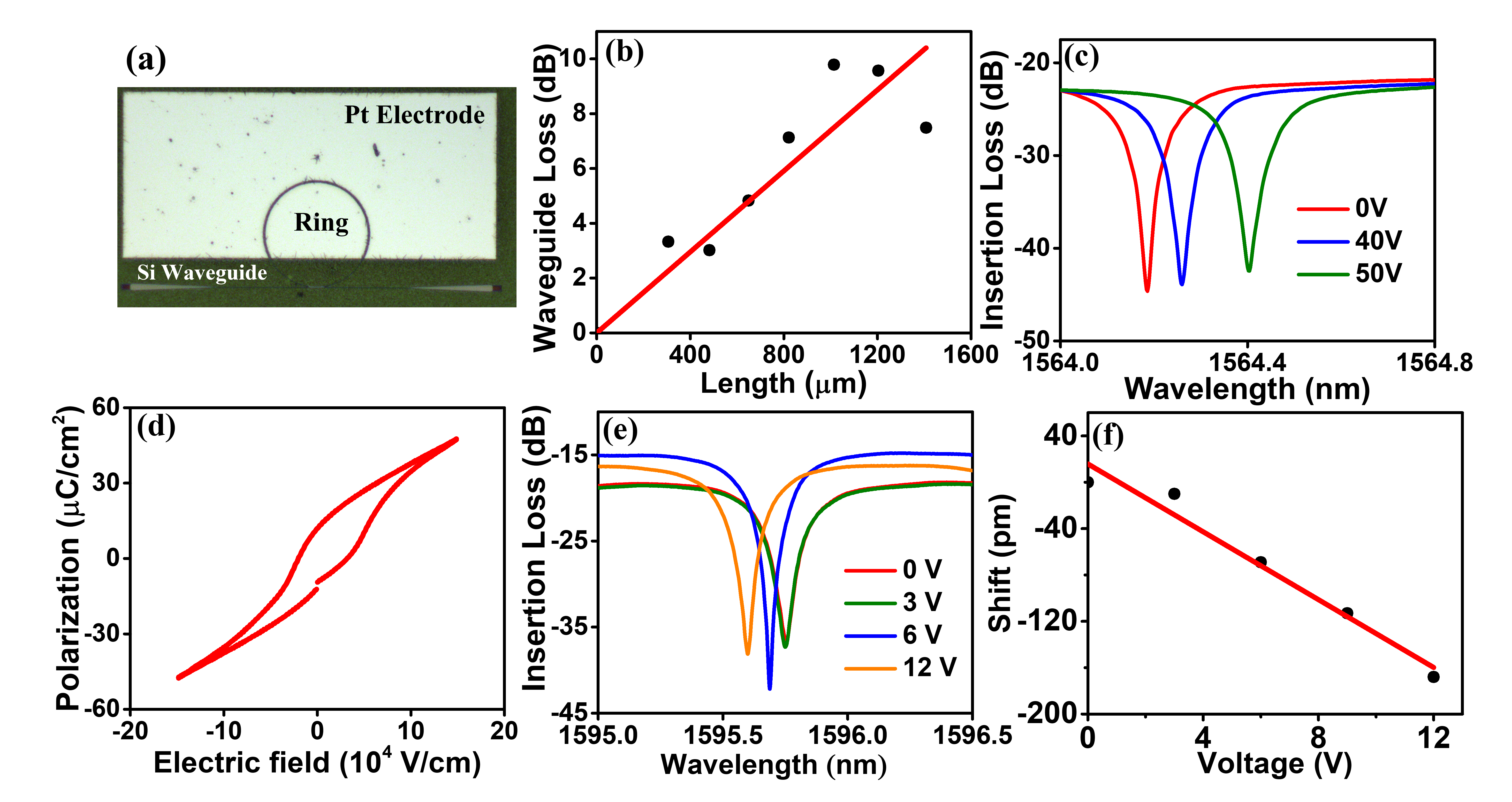}
\caption{(a) Fabricated Ring resonator-based EO modulator; (b) Waveguide loss measured for varying waveguide length ; (c) P-E loop characterization of the device with the co-planar electrode ; (d) EO characterization pre-poling and (e) EO characterization post-poling, and (f) shift in optical spectra variation with applied voltage.}
\label{fig:EO_char}
\end{figure}

The set-up to characterize the EO response of a micro-ring resonator comprised a tunable laser source, a DC power source, and an optical spectrum analyzer. \textbf{Figure \ref{fig:EO_char}(a)} shows the fabricated Si ring resonator with PZT and Pt electrodes used for the EO response characterization. In addition to the ring resonator, Si waveguides with varying length of PZT deposited on them is used to characterize the propagation loss. An excess propagation loss of $\approx$6.5 dB/mm is measured on the PZT loaded Si waveguide (\textbf{Figure \ref{fig:EO_char}(b)}). The unpolled  spectral response of the ring as a function of the applied voltage is shown in \textbf{Figure \ref{fig:EO_char}(c)}. A linear red shift is observed with the applied field due to the domain alignment or poling, which is expected. Polling is performed on the fabricated device with an applied voltage of 70 V for 1 hr and allowed to cool for another 45 min to avoid the thermal shift in the spectra.  
 
 In \textbf{Figure \ref{fig:EO_char}(d)} the P-E loop characterization of the PZT film post-fabrication process indicating piezo-electric property with a coercive field of $\approx$55 kV/cm that closely matches with the characterized PZT film post-deposition and annealing (\textbf{Figure \ref{fig:Functional_char}(d)}). \textbf{Figure \ref{fig:EO_char}(e)}  show the spectral characteristics of polled ring resonator. A characteristic blue shift is observed that confirms EO response. We observe a response of 14 pm/V (\textbf{Figure  \ref{fig:EO_char}(f)}). The red and blue shift corresponds to the process of domain alignment and the net shift in spectra corresponds to the degree of domain alignment and its stability. The response is solely due to the linear EO effect(Pockels effect) corroborated by the pA leakage current and hence rules out the Joule’s heating and charge effects. The waveguide loss and efficiency of the EO modulator can further be improved by increasing the orientation of the deposited PZT film, improving the etch recipe of the PZT film, and improving the architecture of the EO device.

\section{Conclusion}
 
In summary, we have presented a detailed study and optimization process of a sputtered deposited PZT film on an SOI platform for photonic applications. Screening of MgO, TiO$_{2} $ and Pt as a buffer layer is performed to achieve a continuous film for PZT growth. We report that a rutile phase of TiO$_2$ is preferred as a buffer layer to achieve desired PZT phase. However, the film crack density is higher in TiO$_{2} $ buffered layer. While MgO and Pt through optimisation of the annealing temperature and ramp rate, we demonstrate desired PZT phase with low crack density.

The optimized PZT on MgO, TiO$_{2} $ and Pt had a surface roughness of $\approx$2 nm, $\approx$6 nm and $\approx$10 nm, respectively. Electrical characterization of the film on Pt and MgO buffer layer yielded a split peak in the C-V curve and P-E loop with a peak capacitance and polarization of 3 nF,40 $\mu$C/cm$^2$ and 1.8 nF and 55 $\mu$C/cm$^2$ for co-planar electrode respectively. We demonstrate an EO modulator based on MgO-buffered PZT film integration on Si. We obtained a measured peak polarization of $\approx$60 $\mu$C/cm$^2$ with a coercive field of $\approx$ 55 kV/cm. An electro-optic response of 14 pm/V is achieved with a sputter-deposited PZT-on-Si micro-ring resonator. To our knowledge, this is the first demonstration of an electro-optic response from a sputter-deposited PZT.

\section*{Acknowledgments}
SKS thanks Professor Ramakrishna Rao chair fellowship.

\bibliographystyle{unsrt}  
\bibliography{biblio_file}

\begin{thebibliography}{10}

\bibitem{Multi-photon}
Ting Wang, Nalla Venkatram, Jacek Gosciniak, Yuanjing Cui, Guodong Qian, Wei
  Ji, and Dawn T.~H. Tan.
\newblock Multi-photon absorption and third-order nonlinearity in silicon at
  mid-infrared wavelengths.
\newblock {\em Opt. Express}, 21(26):32192--32198, Dec 2013.

\bibitem{chen2023review}
Xiaojun Chen, Jiao Lin, and Ke~Wang.
\newblock A review of silicon-based integrated optical switches.
\newblock {\em Laser \& Photonics Reviews}, page 2200571, 2023.

\bibitem{opt4010003}
Arianna Elefante, Stefano Dello~Russo, Fabrizio Sgobba, Luigi Santamaria~Amato,
  Deborah~Katia Pallotti, Daniele Dequal, and Mario Siciliani~de Cumis.
\newblock Recent progress in short and mid-infrared single-photon generation: A
  review.
\newblock {\em Optics}, 4(1):13--38, 2023.

\bibitem{Perspective}
Bingcheng Pan, Hongxuan Liu, Yishu Huang, Zejie Yu, Huan Li, Yaocheng Shi, Liu
  Liu, and Daoxin Dai.
\newblock Perspective on lithium-niobate-on-insulator photonics utilizing the
  electro-optic and acousto-optic effects.
\newblock {\em ACS Photonics}, 0(0):null, 0.

\bibitem{Ban:22}
Dasai Ban, Guolei Liu, Hongyan Yu, Yingchun Wu, and Feng Qiu.
\newblock Low driving voltage and low optical loss electro-optic modulators
  based on lead zirconate titanate thin film on silicon substrate.
\newblock {\em J. Lightwave Technol.}, 40(9):2939--2943, May 2022.

\bibitem{BaTiO3}
Felix Eltes, Christian Mai, Daniele Caimi, Marcel Kroh, Youri Popoff, Georg
  Winzer, Despoina Petousi, Stefan Lischke, J.~Elliott Ortmann, Lukas
  Czornomaz, Lars Zimmermann, Jean Fompeyrine, and Stefan Abel.
\newblock A batio3-based electro-optic pockels modulator monolithically
  integrated on an advanced silicon photonics platform.
\newblock {\em Journal of Lightwave Technology}, 37(5):1456--1462, 2019.

\bibitem{du1998crystal}
Xiao-hong Du, Jiehui Zheng, Uma Belegundu, and Kenji Uchino.
\newblock Crystal orientation dependence of piezoelectric properties of lead
  zirconate titanate near the morphotropic phase boundary.
\newblock {\em Applied physics letters}, 72(19):2421--2423, 1998.

\bibitem{feutmba2020strong}
Gilles~Freddy Feutmba, Tessa Van~de Veire, Irfan Ansari, John~P George, Dries
  Van~Thourhout, and Jeroen Beeckman.
\newblock A strong pockels pzt/si modulator for efficient electro-optic tuning.
\newblock In {\em Integrated Photonics Research, Silicon and Nanophotonics},
  pages ITu1A--6. Optica Publishing Group, 2020.

\bibitem{spirin1998measurement}
Vasilii~V Spirin, Changho Lee, and Kwangsoo No.
\newblock Measurement of the pockels coefficient of lead zirconate titanate
  thin films by a two-beam polarization interferometer with a reflection
  configuration.
\newblock {\em JOSA B}, 15(7):1940--1946, 1998.

\bibitem{uchiyama2007electro}
Kiyoshi Uchiyama, Atsushi Kasamatsu, Yohei Otani, and Tadashi Shiosaki.
\newblock Electro-optic properties of lanthanum-modified lead zirconate
  titanate thin films epitaxially grown by the advanced sol--gel method.
\newblock {\em Japanese journal of applied physics}, 46(3L):L244, 2007.

\bibitem{haertling1972recent}
GH~Haertling and CE~Land.
\newblock Recent improvements in the optical and electrooptic properties of
  plzt ceramics.
\newblock {\em Ferroelectrics}, 3(1):269--280, 1972.

\bibitem{chen1989anionic}
Chuangtian Chen, Yicheng Wu, and Rukang Li.
\newblock The anionic group theory of the non-linear optical effect and its
  applications in the development of new high-quality nlo crystals in the
  borate series.
\newblock {\em International Reviews in Physical Chemistry}, 8(1):65--91, 1989.

\bibitem{SAHOO2016299}
M.P.K. Sahoo, Yajun Zhang, and Jie Wang.
\newblock Enhancement of ferroelectric polarization in layered bazro3/batio3
  superlattices.
\newblock {\em Physics Letters A}, 380(1):299--303, 2016.

\bibitem{NUFFER20003783}
J~Nuffer, D.C Lupascu, and J~Rödel.
\newblock Damage evolution in ferroelectric pzt induced by bipolar electric
  cycling.
\newblock {\em Acta Materialia}, 48(14):3783--3794, 2000.

\bibitem{singh2021sputter}
Suraj Singh and Shankar~Kumar Selvaraja.
\newblock Sputter-deposited pzt-on-silicon electro-optic modulator.
\newblock In {\em 2021 IEEE Photonics Conference (IPC)}, pages 1--2. IEEE,
  2021.

\bibitem{ZHU201922324}
Minmin Zhu, Haizhong Zhang, Zehui Du, and Chongyang Liu.
\newblock Structural insight into the optical and electro-optic properties of
  lead zirconate titanate for high-performance photonic devices.
\newblock {\em Ceramics International}, 45(17, Part A):22324--22330, 2019.

\bibitem{sakashita1991preparation}
Yukio Sakashita, Toshiyuki Ono, Hideo Segawa, Kouji Tominaga, and Masaru Okada.
\newblock Preparation and electrical properties of mocvd-deposited pzt thin
  films.
\newblock {\em Journal of applied physics}, 69(12):8352--8357, 1991.

\bibitem{das2001preparation}
RN~Das, A~Pathak, SK~Saha, S~Sannigrahi, and P~Pramanik.
\newblock Preparation, characterization and property of fine pzt powders from
  the poly vinyl alcohol evaporation route.
\newblock {\em Materials research bulletin}, 36(9):1539--1549, 2001.

\bibitem{izyumskaya2006growth}
Natalia Izyumskaya, Vitaliy Avrutin, Xing Gu, Umit Ozgur, Bo~Xiao, Tae~Dong
  Kang, Hosun Lee, and Hadis Morkoc.
\newblock Growth of high-quality pb (zrxti1-x) o3 films by peroxide mbe and
  their optical and structural characteristics.
\newblock {\em MRS Online Proceedings Library (OPL)}, 966, 2006.

\bibitem{alexander2018nanophotonic}
Koen Alexander, John~P George, Jochem Verbist, Kristiaan Neyts, Bart Kuyken,
  Dries Van~Thourhout, and Jeroen Beeckman.
\newblock Nanophotonic pockels modulators on a silicon nitride platform.
\newblock {\em Nature communications}, 9(1):1--6, 2018.

\bibitem{busch1992linear}
JR~Busch, SD~Ramamurthi, SL~Swartz, and VE~Wood.
\newblock Linear electro-optic response in sol-gel pzt planar waveguides.
\newblock {\em Electronics Letters}, 28(17):1591--1592, 1992.

\bibitem{nakada2009lanthanum}
Masafumi Nakada, Takanori Shimizu, Hiroshi Miyazaki, Hiroki Tsuda, Jun Akedo,
  and Keishi Ohashi.
\newblock Lanthanum-modified lead zirconate titanate electro-optic modulators
  fabricated using aerosol deposition for lsi interconnects.
\newblock {\em Japanese Journal of Applied Physics}, 48(9S1):09KA06, 2009.

\bibitem{feutmba2019hybrid}
Gilles~F Feutmba, John~P George, Koen Alexander, Dries Van~Thourhout, and
  Jeroen Beeckman.
\newblock Hybrid pzt/si tm/te electro-optic phase modulators.
\newblock In {\em Integrated Optics: Devices, Materials, and Technologies
  XXIII}, volume 10921, pages 85--91. SPIE, 2019.

\bibitem{lee1999drying}
Changho Lee, Vasili Spirin, Hanwook Song, and Kwangsoo No.
\newblock Drying temperature effects on microstructure, electrical properties
  and electro-optic coefficients of sol-gel derived pzt thin films.
\newblock {\em Thin Solid Films}, 340(1-2):242--249, 1999.

\bibitem{swartz1991sol}
SL~Swartz, SD~Ramamurthi, JR~Busch, and VE~Wood.
\newblock Sol-gel pzt films for optical waveguides.
\newblock {\em MRS Online Proceedings Library (OPL)}, 243:533, 1991.

\bibitem{ramamurthi1992electrical}
SD~Ramamurthi, SL~Swartz, KR~Marken, JR~Busch, and VE~Wood Battelle.
\newblock Electrical and optical properties of sol-gel processed pb (zr, ti) o3
  films.
\newblock {\em MRS Online Proceedings Library (OPL)}, 271, 1992.

\bibitem{teowee1995electro}
G~Teowee, JT~Simpson, Tianji Zhao, M~Mansuripur, JM~Boulton, and DR~Uhlmann.
\newblock Electro-optic properties of sol-gel derived pzt and plzt thin films.
\newblock {\em Microelectronic engineering}, 29(1-4):327--330, 1995.

\bibitem{muralt2006texture}
Paul Muralt.
\newblock Texture control and seeded nucleation of nanosize structures of
  ferroelectric thin films.
\newblock {\em Journal of applied physics}, 100(5):051605, 2006.

\bibitem{fushimi1967phase}
Shoichi Fushimi and Takuro Ikeda.
\newblock Phase equilibrium in the system pbo-tio2--zro2.
\newblock {\em Journal of the American Ceramic Society}, 50(3):129--132, 1967.

\bibitem{itoh1991mocvd}
H~Itoh, K~Kashihara, T~Okudaira, K~Tsukamoto, and Y~Akasaka.
\newblock Mocvd for pzt thin films by using novel metalorganic sources.
\newblock In {\em International Electron Devices Meeting 1991 [Technical
  Digest]}, pages 831--834. IEEE, 1991.

\bibitem{zhang2019flexible}
Shaohui Zhang, Long Zhang, Lujia Wang, Fengxia Wang, and Gebo Pan.
\newblock A flexible e-skin based on micro-structured pzt thin films prepared
  via a low-temperature pld method.
\newblock {\em Journal of Materials Chemistry C}, 7(16):4760--4769, 2019.

\bibitem{abe1991pzt}
Kazuhide Abe, Hiroshi Tomita, Hiroshi Toyoda, Motomasa Imai~Motomasa Imai, and
  Yukari Yokote~Yukari Yokote.
\newblock Pzt thin film preparation on pt-ti electrode by rf sputtering.
\newblock {\em Japanese journal of applied physics}, 30(9S):2152, 1991.

\bibitem{kim1997influence}
Seung-Hyun Kim, Chang-Eun Kim, and Young-Jei Oh.
\newblock Influence of al2o3 diffusion barrier and pbtio3 seed layer on
  microstructural and ferroelectric characteristics of pzt thin films by
  sol-gel spin coating method.
\newblock {\em Thin Solid Films}, 305(1-2):321--326, 1997.

\bibitem{basit1998growth}
Nasir~Abdul Basit, Hong~Koo Kim, and Jean Blachere.
\newblock Growth of highly oriented pb (zr, ti) o 3 films on mgo-buffered
  oxidized si substrates and its application to ferroelectric nonvolatile
  memory field-effect transistors.
\newblock {\em Applied physics letters}, 73(26):3941--3943, 1998.

\bibitem{song1997effects}
Han~Wook Song, Joon~Sung Lee, Dae-Weon Kim, Kwang~Ho Kim, Tae-Hyun Sung, and
  Kwangsoo No.
\newblock Effects of buffer layer on the fabrication and characteristics of
  ferroelectric thin films.
\newblock {\em MRS Online Proceedings Library (OPL)}, 493, 1997.

\bibitem{yong2000effects}
Zhu Yong-Fa, Yan Pei-Yu, Yi~Tao, Cao Li-Li, and Li~Long-Tu.
\newblock Effects of pt diffusion barrier layer on the interface reaction and
  electric properties of pzt film/si (111) sample.
\newblock {\em Chinese Journal of Chemistry}, 18(3):328--334, 2000.

\bibitem{choi2015investigation}
Sujin Choi, Juyun Park, Sung-Wi Koh, and Yong-Cheol Kang.
\newblock Investigation of post annealing effect on the pzt thin films.
\newblock {\em Journal of the Chosun Natural Science}, 8(4):244--249, 2015.

\bibitem{willems1995influence}
Geert~J Willems, Dirk~J Wouters, and Herman~E Maes.
\newblock Influence of the pt electrode on the properties of sol-gel pzt-films.
\newblock {\em Microelectronic Engineering}, 29(1-4):217--220, 1995.

\bibitem{kukushkin2012mechanism}
SA~Kukushkin, I~Yu Tentilova, and IP~Pronin.
\newblock Mechanism of the phase transformation of the pyrochlore phase into
  the perovskite phase in lead zirconate titanate films on silicon substrates.
\newblock {\em Physics of the Solid State}, 54:611--616, 2012.

\bibitem{peng2013improvement}
QX~Peng, CG~Wu, WB~Luo, L~Jin, WL~Zhang, C~Chen, and XY~Sun.
\newblock The improvement of pyroelectric properties of pzt thick films on si
  substrate by tiox barrier layer.
\newblock {\em Infrared Physics \& Technology}, 58:51--55, 2013.

\bibitem{gilmore2003growth}
Walter~M Gilmore, Soma Chattopadhyay, Alex Kvit, AK~Sharma, Clinton~B Lee,
  Ward~J Collis, Jagannathan Sankar, and J~Narayan.
\newblock Growth, characterization, and electrical properties of pbzr0. 52ti0.
  48o3 thin films on buffered silicon substrates using pulsed laser deposition.
\newblock {\em Journal of materials research}, 18(1):111--114, 2003.

\bibitem{tonejc2002analysis}
AM~Tonejc, I~Djerdj, and A~Tonejc.
\newblock An analysis of evolution of grain size-lattice parameters dependence
  in nanocrystalline tio2 anatase.
\newblock {\em Materials Science and Engineering: C}, 19(1-2):85--89, 2002.

\bibitem{yano1994epitaxial}
Y~Yano, K~Iijima, Y~Daitoh, T~Terashima, Y~Bando, Y~Watanabe, H~Kasatani, and
  H~Terauchi.
\newblock Epitaxial growth and dielectric properties of batio3 films on pt
  electrodes by reactive evaporation.
\newblock {\em Journal of Applied Physics}, 76(12):7833--7838, 1994.

\bibitem{sreemany2009influence}
Monjoy Sreemany, Ankita Bose, and Suchitra Sen.
\newblock Influence of chemical composition, phase and thickness of tiox
  (x$\leq$ 2) seed layer on the growth and orientation of the perovskite phase
  in sputtered pzt thin films.
\newblock {\em Materials Chemistry and Physics}, 115(1):453--462, 2009.

\end{thebibliography}

\end{document}